\documentclass[twocolumn,pre,float]{revtex4-1}
\usepackage{graphicx}
\usepackage{pgfplots,xcolor}
\usepackage{epstopdf}
\usepackage{amsmath,amsfonts,amssymb,paralist,mathrsfs}
\usepackage{bm}
\usepackage{verbatim}
\usepackage{enumerate}
\usepackage{natbib}
\begin{document}

\title{Transport and fluctuation--dissipation relations in asymptotic and 
pre--asymptotic diffusion across channels with variable section}

\author{Giuseppe~Forte$^{1}$, 
Fabio~Cecconi$^{2}$, 
Angelo~Vulpiani$^{1,3}$}
\affiliation{$^1$ Dipartimento di Fisica Universit\`a di 
Roma ``Sapienza'', 
Piazzale Aldo Moro 2, I--00185 Roma, Italy\\
$^2$ CNR--Istituto dei Sistemi Complessi (ISC), Via dei Taurini 19, 
I--00185 Roma, Italy\\
$^3$ CNR--Istituto dei Sistemi Complessi (ISC), Piazzale Aldo Moro 2, 
I--00185 Roma, Italy.}

\begin{abstract}
We study the asymptotic and pre--asymptotic diffusive properties of Brownian 
particles in channels whose section varies periodically in space. 
The effective diffusion coefficient $D_{\mathrm{eff}}$ is numerically 
determined by the asymptotic behavior of the root mean square displacement
in different geometries, considering even cases of steep
variations of the channel boundaries. 
Moreover, we compared the numerical results to the predictions from
the various corrections proposed in the literature to the well 
known Fick-Jacobs approximation.
Building an effective one dimensional equation for the longitudinal 
diffusion, we  obtain  an approximation for the effective diffusion 
coefficient. Such a result goes beyond a perturbation approach, and it is 
in good agreement with the actual values obtained by the numerical simulations. 
We discuss also the pre--asymptotic diffusion which is observed 
up to a crossover time whose value, in the presence of strong spatial 
variation of the channel cross section, can be very large.
In addition, we show how the Einstein's relation between the mean drift 
induced by a small external field and the mean square displacement of the 
unperturbed system is valid in both asymptotic and pre--asymptotic 
regimes.
\end{abstract} 

\maketitle

\section*{Introduction}
The diffusive transport across non homogeneous channels 
\cite{RR_01,Hanggi1_07,Hanggi4_08,Mar_3,Mar_5} 
is one of the most interesting examples of dynamics 
influenced by the geometrical properties of the surrounding environment. 
Many important phenomena and applications 
are related to constrained diffusion such as, the flow in 
porous materials \cite{Dullien91porous,Rolando_05}, the separation
of DNA fragments moving in narrow channels \cite{Han00,Heng04}, the 
emergence of a pre--asymptotic subdiffusive transport 
in spiny dendrites \cite{Santamaria} and nuclear magnetic resonance 
measurements in tissues of complex morphology 
\cite{Grebenkov07_NMR,Sen04}.

An interesting feature is the slowdown of the diffusion due   
the trapping mechanism of molecules within compartments 
and dead-end regions offering the possibility of a geometrical 
control of transport rates.

The most common theoretical approach involves the Fick--Jacobs 
(FJ) approximation \cite{Jacobs} and its generalizations 
\cite{Mar_3,GoodKnight}. This method, mainly applicable to 
structures with strong lateral confinement,  
amounts to a dimensional reduction where  
the diffusion within two-- or three--dimensional channels is projected 
onto their centerline, obtaining processes which obey an effective 
one--dimensional diffusion equation.
The validity of the FJ approach in the unbiased 
\cite{Mar_3,Hanggi1_07,Marc5_09} and biased \cite{Mar_5} 
case was extensively studied. 
In particular, one of the main questions related to the unbiased 
diffusion in the asymptotic regime is the derivation of the effective 
diffusion coefficient $D_{\mathrm{eff}}$ along the longitudinal 
direction, as a function of the external geometrical parameters 
\cite{Mar_3}.  
The constant $D_{\mathrm{eff}}$ controls the rapidity of the mass spreading, 
thus affecting, for example, the rate at which the molecules hit 
certain target regions \cite{Hanggi5_06}. 
In this respect, the theoretical knowledge 
can be crucial to design nano-devices with certain 
desirable transport properties.

In this paper we focus on the properties of
diffusive motion within two--dimensional spatially periodic channels. 
We study the asymptotic as well as the pre--asymptotic 
regime using both analytical and numerical techniques.

Using a Markovian approximation within a coarse-graining procedure,
we derive a simple estimation of $D_{\mathrm{eff}}$
without resorting to the FJ theory. Then we compare the theoretical 
predictions with numerical data from Brownian dynamic simulations 
within the channels and with other known approximations from the 
literature. 
As we discuss in this paper, a derivation of $D_{\mathrm{eff}}$  
that is alternative to the FJ approach is particularly important 
in all the cases where the channel boundaries are multivalued functions 
of the longitudinal coordinate. 

We devote a special attention to relationship between the 
average mean square displacement of the unbiased diffusion across the 
channels and the average drift of the biased diffusion: 
Einstein's fluctuation--dissipation Relation (FDR).  
The analytical expression of the asymptotic nonlinear mobility was 
worked out by several authors \cite{Mar_7,Mar_5}; 
however, it is natural to wonder if the FDR still holds true also in the 
pre--asymptotic (transient) regime. 
We show that a generalized FDR also applies 
to the pre--asymptotic diffusion. 

The paper is organized in the following way. 
In Sec.~\ref{sec:recall}, we recall the principal results and 
approximations used to characterize the diffusion within periodic 
channels. 
In Sec.~\ref{sec:asy}, we derive an analytical formula for
$D_{\mathrm{eff}}$ comparing our analytical results 
with numerical simulations and the FJ approach.
We highlight the benefits and limitations of our approach, showing how we 
can estimate $D_{\mathrm{eff}}$ also when the FJ approach does not apply.
Section~\ref{sec:preasy_new} is devoted to the analysis of 
pre--asymptotic diffusion properties.
In Sec.~\ref{sec:FDR} we study the response to a constant external field 
applied along the channel longitudinal direction, analyzing both the asymptotic and the pre--asymptotic dynamical regimes. 
Finally, Sec.~\ref{sec:conclusion} contains a summary of the main 
results and the conclusions.
The Appendix~\ref{app:A}  shows some technical details.

\section{Recalling the diffusion equations in confined systems.
\label{sec:recall}}
\begin{figure}
\includegraphics[clip=true,keepaspectratio,width=\columnwidth]
{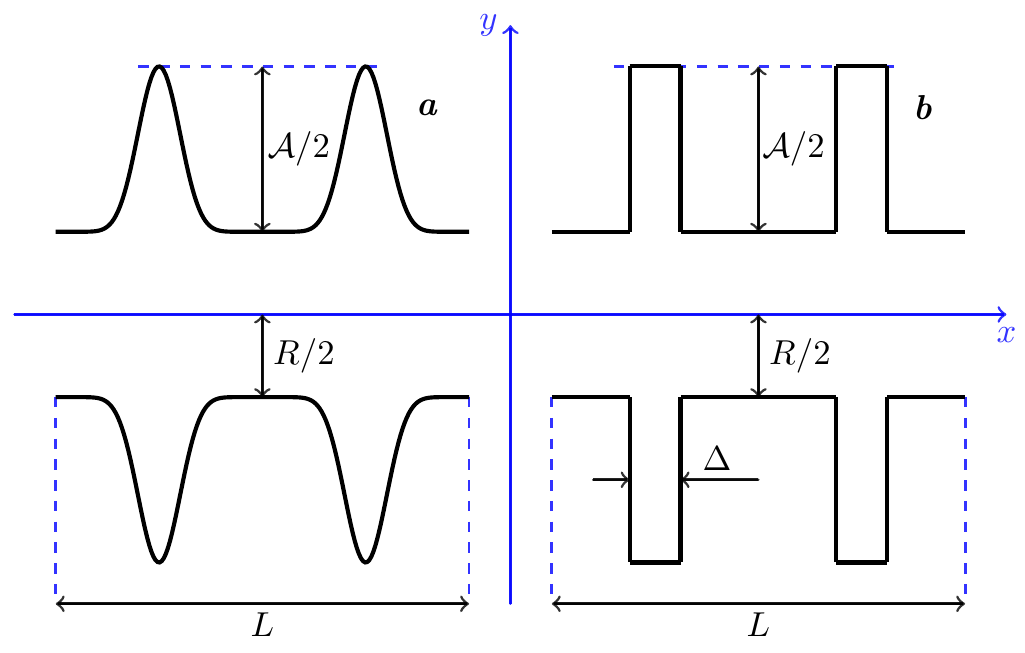}
\caption{Sketch of the periodic structures considered in this work.
(a) Smooth channel (Sm); (b) Sharp channel (Sh).
We identify $H$--regions (humps) of the channel such that 
$R/2<|y|\le \omega(x)$ and the $S$--region (shaft) $|y|\le R/2$.  
Diffusion regimes inside these channels depend on the
ratio $Q = \mathcal{A}/R$ controlling the importance of the $H$ regions 
over the $S$ region. 
When not specified, we consider channels with fixed shaft 
parameters $L=10$, $R=4$, whereas, the size of the $H$--regions 
will be changed to obtain stronger or weaker lateral particle trapping.}
\label{fig:Chann}
\end{figure}

We consider the dynamics of particles in a two--dimensional 
spatially periodic channel (Fig.~\ref{fig:Chann}) 
forming a sufficiently diluted gas well described by the single particle 
approximation. Each particle evolves in the 
presence of a possible external potential $V(\mathbf r)$   
according to the overdamped Langevin equation \citep{Langevin,ZwBook_01}, 
\begin{equation}
\frac{d\mathbf{r}}{dt} = -\frac{\boldsymbol{\nabla}V(\mathbf r)}{\eta}
+\sqrt{\frac{2k_{B}T}{\eta}}\,\boldsymbol{\xi}(t)
\label{eq:lang_over}
\end{equation}
where $k_{B}$ is the Boltzmann's constant, $\eta$ and $T$ are 
respectively the viscous friction coefficient of the fluid filling 
the channel.
The stochastic term $\boldsymbol{\xi}(t)$ is a Gaussian white noise:
$$
\langle \xi^{(i)}(t)\rangle = 0,\quad
\langle \xi^{(i)}(t)\xi^{(j)}(t')\rangle=\delta_{ij}\delta(t-t')
\quad i,j = x,y.
$$
The Fokker--Planck equation \citep{Risken} for the probability density
$\mathscr{P}(\mathbf{r},t)$ corresponding to the stochastic process 
\eqref{eq:lang_over} reads 
\begin{equation}
\begin{cases}
\partial_{t}\mathscr{P}(\mathbf{r},t)+
\boldsymbol{\nabla}\cdot\mathbf{J}(\mathbf{r},t)  =0
\vspace{0.8em}\\
\mathbf{J}(\mathbf{r},t) = -\left[\dfrac{\boldsymbol{\nabla}V(\mathbf{r})
}{\eta}+D_{0}\boldsymbol{\nabla}\right]\mathscr{P}(\mathbf{r},t) \\
\end{cases}
\label{eq:chapFive_stateproblem}
\end{equation}
where $D_{0} = k_{B}T/\eta$ denotes the microscopic diffusion coefficient. 
Moreover, the no--flux boundary conditions have to be imposed to 
take into account the impenetrable nature of the channel walls, 
\begin{equation}
\mathbf{J}(\mathbf{r},t)\cdot\hat{\mathbf{n}}(\mathbf{r}) =0,\qquad 
\mathbf{r}\in\mbox{Channel walls}
\label{eq:boundary}
\end{equation}  
with $\hat{\mathbf{n}}(\mathbf{r})$ being the local unitary normal 
to the channel walls. 
In the following we refer to the case with no external field 
[$V(\mathbf r)=0$]. 
Such a system is trivial only for channels with constant section, for 
which we have the asymptotic behaviour 
$\langle[x(t) - x(0)]^2\rangle =  2 D_0 t$.
When the channel section varies as in Fig.~\ref{fig:Chann}, 
we still expect a large-time standard diffusion, but with a 
renormalized constant,
$$
\langle[x(t) - x(0)]^2\rangle = 2 D_{\mathrm{eff}} t\;,
$$
the value of $D_{\mathrm{eff}}$ is smaller than $D_0$ and depends on the 
variation of the channel section $\sigma(x)$. Therefore one of the 
main issues is determining $D_{\mathrm{eff}}$ once the geometrical 
shape of the channel is assigned. 
Let us assume that the channel is parallel to the $x$--axis and its 
cross section $\sigma(x)$ varies periodically on the longitudinal 
direction. It is convenient to introduce the marginal density 
\citep{KP2_05,Brz3_07,Mar_3}
$\mathscr{G}(x,t)$, defined by
\begin{equation}
\mathscr{G}(x,t) = \int_{-\omega(x)}^{+\omega(x)}
\!\!\!\!\!dy\;\mathscr{P}(x,y,t)
\label{eq:chapFive_marginales}
\end{equation}
where we considered a symmetric channel along 
its longitudinal axis described by the boundary profile 
$y = \pm \omega(x)$, thereof the cross section is $\sigma(x) = 2 \omega(x)$. 

Rather general procedures of the multiscale technique suggest the 
validity at large time of a parabolic equation governing  
the diffusion along the $x$--direction \cite{Jacobs,Zw1_92,KP1_06} 
given by      
\begin{equation}
\frac{\partial\mathscr{G}(x,t)}{\partial t} = \frac{\partial}{\partial x}
\left\{\sigma(x)D(x)\frac{\partial }{\partial x}\left[
\frac{\mathscr{G}(x,t)}{\sigma(x)}\right]\right\}\;.
\label{eq:chapFive_Zw} 
\end{equation} 
Jacobs in his book on diffusive processes \citep{Jacobs} 
used the drastic approximation
\begin{equation}
D_{\mathrm{FJ}}(x)  = D_0\;,
\label{eq:FJDx}
\end{equation}
assuming that in narrow-enough channels the distribution of particles 
in any cross section becomes swiftly uniform (local equilibrium).  

Several authors \cite{Zw1_92,Szabo11} argued that deviations 
from transversal homogeneity are not negligible and need to be taken into 
account by a diffusion coefficient $D=D(x)$ varying with the longitudinal 
coordinate.
The specific expression of $D(x)$ depends on the channel boundaries  
and guessing the appropriate functional form for given geometry  
is a central task of this kind of problems.
Zwanzig for example \citep{Zw1_92} derived a perturbative expression 
for $D(x)$ in a two dimensional channel, 
\begin{equation}
D_{\mathrm{Zw}}(x) =  
D_{0}\left[1+\frac{1}{12}\left(\frac{d\sigma}{dx}\right)^{2}\right]^{-1} 
\label{eq:ZwDx}
\end{equation}
which holds under the hypothesis $|\sigma'(x)|\ll 1$. 

Instead, Reguera and Rub\'i (RR) using a heuristic argument \citep{RR_01} 
proposed the expression
\begin{equation}
D_{\mathrm{RR}}(x) = 
D_{0}\left[1+\frac{1}{4}\left(\frac{d\sigma}{dx}\right)^{2}\right]^{-1/3}\;.  
\label{eq:ZwRR}
\end{equation}

Finally, Kalinay and Percus (KP) performing an elegant perturbative treatment
in order to expand $D(x)$ in $\sigma(x)$ and its derivatives
\citep{KP2_05,KP3_05,KP1_06,KP5_08} obtained, for a 2D channel,
to the lowest order the expression
\begin{equation}
D_{\mathrm{KP}}(x)= 
D_{0}\frac{\arctan{\left(\dfrac{1}{2}
\dfrac{d\sigma(x)}{dx}\right)}}{\dfrac{1}{2}\dfrac{
d\sigma(x)}{dx}}\,.
\label{eq:chapFive_KP}
\end{equation}

It is easily to verify by a series expansion in $|\sigma'(x)|$ 
that, to the lowest order,  all the functional forms coincide 
$D_{\mathrm{KP}}(x) = 
D_{\mathrm{RR}}(x) = 
D_{\mathrm{Zw}}(x) \approx D_0(1-\sigma'(x)^2/12)$.  
 
Once an explicit expression of $D(x)$ has been established, 
Eq.~\eqref{eq:chapFive_Zw} provides the values of $D_{\mathrm{eff}}$ 
through the Lifson--Jackson (LJ) formula \citep{LJ62}
\begin{equation}
D_{\mathrm{eff}}=\frac{1}{\langle\sigma(x)\rangle\left\langle
\displaystyle\frac{1}{D(x)\sigma(x)}\right\rangle}
\label{eq:chapFive_L_J}
\end{equation}
where the angular brackets denote the spatial average over a period 
$L$ of the channel 
$$
\langle f(x)\rangle = \frac{1}{L}\int_{x_{0}}^{x_{0}+L}
\!\!\!\!\!\!\!dx\;f(x),
$$
with $f(x) = f(x+L)$.

Once the asymptotic diffusion process  
has been reduced to the effective one-dimensional PDE 
(\ref{eq:chapFive_Zw}), the problem can be also 
recast into the corresponding one-dimensional Langevin 
equation 
$$
\frac{dx}{dt} = -\frac{dV(x)}{dx} + \sqrt{2D(x)}\xi(t)
$$
with an effective potential
$$
V(x) = -k_{B}T\ln{\sigma(x)}
$$
where $V(x)$ is a periodic function depending on $\sigma(x)$
and represents an entropic potential \cite{Hanggi5_06}.

The perturbative approach of Refs.~\cite{Zw1_92,KP5_08,Marte011} 
fails to describe the most interesting cases where $\mathcal{A} \gtrsim R,L$ 
or becomes particularly involved \cite{KP_abrupt} when dealing with    
not differentiable boundaries of the channel (see Fig.~\ref{fig:Chann}).

\section{Asymptotic Diffusion}
\label{sec:asy}
Let us introduce a simple argument to derive an approximation 
to the effective coefficient $D_{\mathrm{eff}}$ of the
longitudinal diffusion.
 
Of course, the trapping mechanism due to the humps implies that 
$D_{\mathrm{eff}} < D_0$. 
With reference to Fig.~\ref{fig:Chann}, we shall dub 
``Humps'' ($H$) and ``Shaft'' ($S$) the sets of the channel 
such that 
\begin{eqnarray}
H & = & \{(x,y) \in\mathbb{R}^2~|~ R/2 <|y|\le \omega(x)\} \nonumber \\  
S & = & \{(x,y) \in\mathbb{R}^2~|~ |y|\le R/2\}. 
\label{eq:partition}  
\end{eqnarray}
Within the $H$ region the particles spend a certain amount of time before 
coming back to the $S$ region where they contribute to the transport 
along the $x$--direction.

We consider two types of structures sketched in Fig.~\ref{fig:Chann}.
The first (Fig.~\ref{fig:Chann}a) is defined by the smooth 
boundary
\begin{equation}
\omega_{\mathrm{Sm}}(x)=  
\dfrac{R + s(\gamma) \mathcal{A}}{2}+
\dfrac{\mathcal{A}}{2}\sin^{\gamma}\left(\dfrac{2\pi x}{L}\right)
\label{eq:Sin_Chann_w}
\end{equation}
where $\gamma$ is an integer tuning the shape of the
humps: the larger $\gamma$, the sharper the sinusoidal humps and   
$s(\gamma=2n)=0$, $s(\gamma=2n+1)=1$. 
The extra term, $s(\gamma)\mathcal{A}/2$, when $\gamma$ turns from even
to odd values is necessary to avoid the upper- and lower-boundary 
overlap and to keep  the $S$-region width fixed to $R$. 
Analogously, the period of the ``even'' and ``odd'' profiles changes 
from $L/2$ to $L$. 
Hereafter, this channel is referred to as the Smooth channel (Sm). 

The other structure that we name Sharp channel (Sh), 
Fig.~\ref{fig:Chann}b, is characterized by the step-like profile
\begin{equation}
\omega_{\mathrm{Sh}}(x)=
\begin{cases} 
\dfrac{R+\mathcal{A}}{2}   &  
      \dfrac{L}{4} - \dfrac{\Delta}{2} \le 
|x| < \dfrac{L}{4} + \dfrac{\Delta}{2} 
\vspace{0.8em} \\ 
\dfrac{R}{2}  &  \mbox{elsewhere}
\end{cases}
\label{eq:Rec_Chann_w}
\end{equation}
The period of this channel is $L/2$. 
More complicated boundaries will be taken into account in the following. 
Unless otherwise stated, throughout the text and in all simulations, 
we consider channels with fixed shaft parameters: $L=10$, $R=4$,
whereas, the parameters controlling the size and the shape 
of the $H$-regions, such as $\mathcal{A}$, $\gamma$ and $\Delta$,  
will be tuned to obtain different degrees of dead-end trapping.

We introduce a phenomenological Markov process to estimate 
$D_{\mathrm{eff}}$. 
Let us observe that in the long time limit, the motion along the 
transverse direction becomes stationary.  
Thus the probabilities $P_H(t)$ and $P_S(t) = 1-P_H(t)$ to be 
in the $H$ and $S$ regions respectively become constant at large times, 
\begin{equation}
\lim_{t\to\infty} P_{S}(t) = P^{eq}_{S} = \frac{\mu(S)}{\mu(H)+\mu(S)},
\label{eq:PT}
\end{equation} 
where $\mu(S)$ and $\mu(H)$ are the areas (measure) of the $S$ and $H$ 
regions respectively within a single period of the channel. 
Since we are interested only in the longitudinal transport process, 
we can approximate the particle motion as a random walk 
on a 1D ``lattice'' where the walkers can either jump to the left or to the 
right with probability $\mu(S)$ or they can sit on the same site with 
probability $\mu(H)$.     
In this coarse--grained representation which mimics the 
longitudinal diffusion process, $D_{\mathrm{eff}}$ can be expressed 
in the form 
\begin{equation}
D_{\mathrm{eff}}=D_{0}P_S^{eq}\;.
\label{eq:Deff}
\end{equation}
Notice that the result coincides with the one proposed in 
Refs.~\cite{Brz10_07,Berez_2011}.

Equations~\eqref{eq:PT} and~\eqref{eq:Deff} when applied to the 
Sm-channel \eqref{eq:Sin_Chann_w} provide the result
\begin{equation}
\frac{D_{\mathrm{eff}}}{D_0} = 
\begin{cases}
\dfrac{R}{R + \dfrac{1}{\pi} B\left(\dfrac{1}{2},\dfrac{\gamma + 1}{2}\right)\mathcal{A}} & \mbox{if~} \gamma = 2n 
\vspace{0.5em}\\
\dfrac{R}{R + \mathcal{A}} & \mbox{if~} \gamma = 2n+1
\end{cases}
\label{eq:sinDeff}
\end{equation}
where $B(a,b) = \Gamma(a)\Gamma(b)/\Gamma(a+b)$ is the 
Euler's $\beta$ function. 

The dependence of $D_{\mathrm{eff}}$ on even $\gamma$'s  
at $R$ and $\mathcal{A}$ fixed is 
determined by the behavior of the function $B[1/2,(\gamma + 1)/2]$, 
which monotonically decreases with $\gamma$ and asymptotically 
scales like $B \sim \sqrt{2/(\pi\gamma)}$. 
This implies that $D_{\mathrm{eff}}$ approaches $D_0$ at large even 
$\gamma$s, which is not a surprising situation if one considers that 
a very large $\gamma$ determines inaccessible narrow Humps in the 
profile \eqref{eq:Sin_Chann_w},
consequently, the particles are constrained to perform a free diffusion 
along the Shaft region.    

For odd $\gamma$, the area cancellation in $\mu(H)$ yields a
result that is independent of the Hump's shape and the 
$\gamma$--value as well.
This peculiarity makes the applicability of the areas' formula 
critical to ``odd'' channels, as we discuss later.
 
For the Sharp channel \eqref{eq:Rec_Chann_w}, we obtain 
$$
\mu(S) = RL, \qquad \mu(H) = 2\mathcal{A}\Delta\;,
$$
and the effective coefficient reads
\begin{equation}
\frac{D_{\mathrm{eff}}}{D_{0}} = \frac{R L}{R L + 2\mathcal{A}\Delta}.
\label{eq:rectDeff}
\end{equation}
Equations \eqref{eq:sinDeff} and \eqref{eq:rectDeff} can be recast in a 
common form highlighting the relevant dependence on the ratio 
$Q = \mathcal{A}/R$,      
\begin{equation}
\frac{D_{\mathrm{eff}}}{D_{0}} = \dfrac{1}{1 + g_H Q};
\label{eq:genDeff}
\end{equation}
where $g_H$ is a coefficient depending on the finer details of the 
hump regions; in particular, $g_H = 1$ ($\gamma$ odd), $g_H= B[1/2,(\gamma+1)/2]/\pi$ ($\gamma$ even) for a Sm-channel, 
while $g_H = 2\Delta/L$ in a Sh-channel. 

It is interesting to remark that Eq.\eqref{eq:rectDeff} is amenable to an 
alternative derivation suggested by the general multiscale techniques 
\cite{weinanE_book} where the concept of an effective diffusion equation 
along the channel centerline naturally arises, namely
\begin{equation}
\partial_t P(x,t) = \partial_x\{\tilde{D}(x) \partial_x P(x,t)\}
\label{eq:simple}
\end{equation}
note that now $\tilde{D}(x)$ embodies all the inhomogeneities of the 
problem and thus does not coincide with $D(x)$ of 
Eq.~\eqref{eq:chapFive_Zw}. 
Equations \eqref{eq:chapFive_Zw} and \eqref{eq:simple} need not be related
in a direct mathematical way as they  provide only
an equivalent mesoscopic representation of the same asymptotic
process: the large time diffusive behavior when the transversal
homogenization has been achieved. 

The above formulation has the advantage that the result 
\eqref{eq:rectDeff} can be still recovered  through 
\eqref{eq:chapFive_L_J} providing one assumes a local diffusion 
coefficient such that  
\begin{equation}
\tilde{D}(x) =
\begin{cases} 
D_0 \dfrac{R}{\mathcal{A}+ R} &  
      \dfrac{L}{4} - \dfrac{\Delta}{2} \leq
|x| < \dfrac{L}{4} + \dfrac{\Delta}{2} 
\vspace{0.8em}\\ 
D_0                            & \mbox{elsewhere}
\end{cases}.
\label{eq:Djump}
\end{equation}
Such a proposal can be rationalized as follows. A particle 
can fully contribute to the diffusion only when its $y$ coordinate 
lies in the $S$-region. 
Accordingly, if the particle is in the $S$-region, 
its diffusion occurs with a free coefficient $D_0$, whereas in the $H$-region 
the diffusion coefficient is depressed by a factor $R/(\mathcal{A} + R)$. 
Now from 
expression \eqref{eq:Djump} we can compute the 
effective diffusion coefficient as
$$
D_{\mathrm{eff}} = 
\dfrac{1}{\left\langle\dfrac{1}{\tilde{D}(x)}\right\rangle}\,.
$$

This approach, in a philosophy similar to the homogenization technique, 
amounts to considering a flat channel while shifting 
its inhomogeneity to the diffusion coefficient 
and it has the advantage to be also applicable to not differentiable 
boundary profiles.     

In order to test the quality of the approximation~\eqref{eq:genDeff}, 
we performed numerical simulations by integrating the Langevin 
equation \eqref{eq:lang_over} with $V(\mathbf{r}) = 0$.  
The no-flux condition has been implemented
both via a simple rejection algorithm and via an elastic-reflection method 
in the collision of a particle against the wall. 
The results were independent of the method and we   
opted for the rejection method which is faster and of easier implementation. 
We simulated $N=7\times 10^{4}$ independent particle trajectories 
via a standard Euler's scheme with $D_{0}=1$ and a time step 
$\Delta t = 0.005\tau$, 
where $\tau = R^2/2D_0$ and $R$ is the fixed cross section of the shaft 
region (see Fig.~\ref{fig:Chann}). 
Hence, for each time step, the mean step length 
$\Delta \ell = \sqrt{2D_0 \Delta t} \simeq 0.07 R$,     
turns out to be reasonably smaller than the typical geometrical 
length of the channel but large enough to avoid excessively long 
simulation runs. 
\begin{figure}
\includegraphics[clip=true,keepaspectratio,width=0.9\columnwidth]
{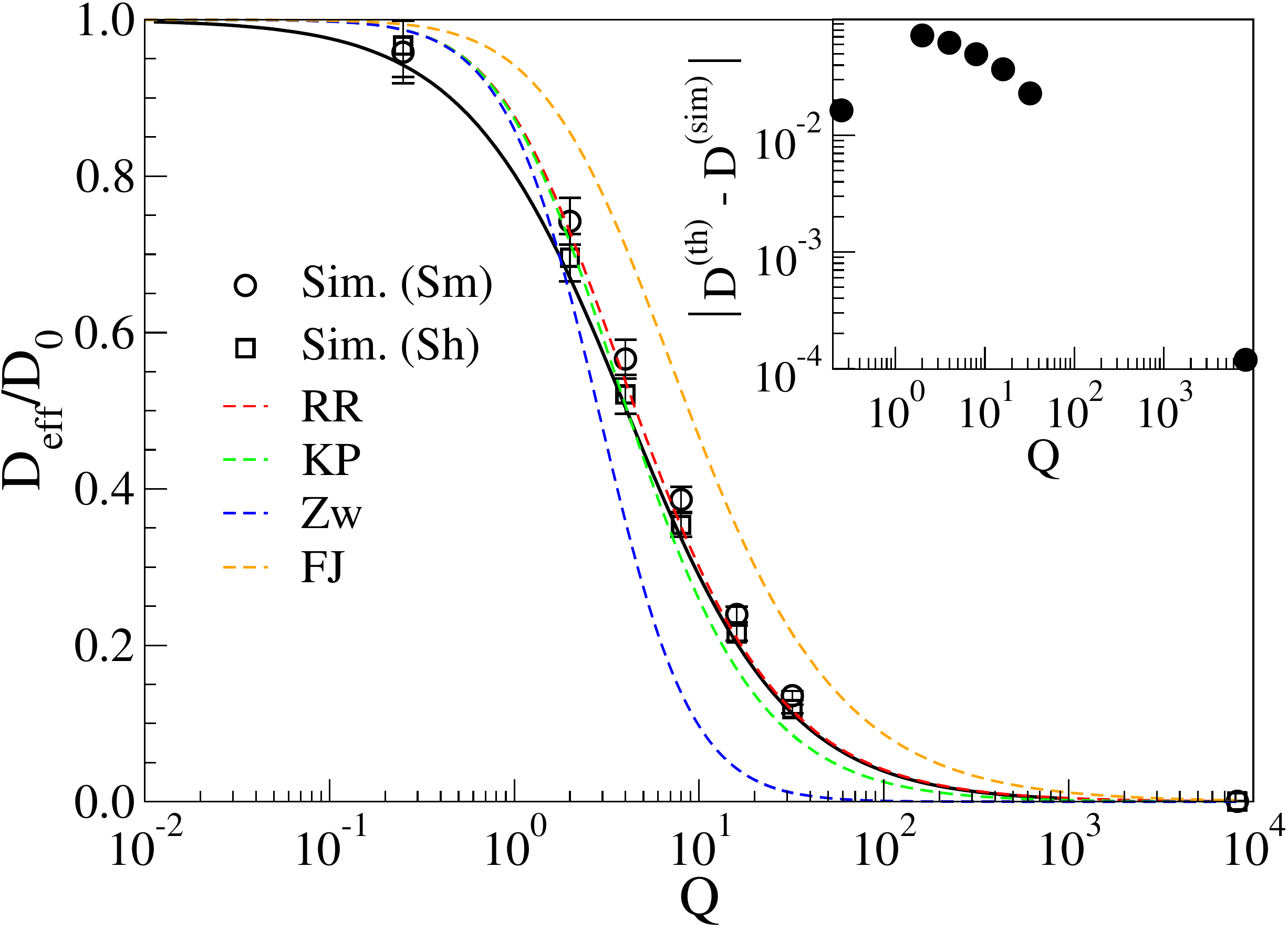}
\caption{Plot of Eq.~\eqref{eq:genDeff}  
vs $Q = \mathcal{A}/R$ (black line) together with the simulated  
$D_{\mathrm{eff}}$ data obtained by the slope of
$\langle\Delta x^2(t)\rangle$ in the linear regime.  
Symbols correspond to values $Q = 2^{-2},2^{1},2^{3},\cdots 2^{13}$;    
circles refer to the Sm-channel with $L = 10, R = 4, \gamma=10$, while 
squares refer to the Sh-channel with $L = 10, R = 4, 
\Delta = 1.23$. 
Since the channels are equivalent (same area), the data from both 
structures are supposed to collapse onto each other. 
The percentage error is within 4\%; the {\em inset} instead 
shows the deviation $|D_{\mathrm{eff}}(\mathrm{th}) - D_{\mathrm{eff}}
(\mathrm{sim})|$ of the black line from the open circles 
in the main panel.
For a further comparison, we report the $D_{\mathrm{eff}}$ of the 
Sm-channel computed by Eq.\eqref{eq:chapFive_L_J}
using the local diffusion coefficients mentioned in the text: 
$D_{RR}(x)$ (red), $D_{KP}(x)$ (green), $D_{Zw}(x)$ (blue), and 
$D_{FJ}(x)$ (orange).} 
\label{fig:Deff}
\end{figure}

The numerical results are shown in Fig.~\ref{fig:Deff}, 
where formula~\eqref{eq:genDeff} is compared to the simulation data
obtained from the asymptotic behaviour of the longitudinal MSD
in the Sm- and Sh-channels at different values of 
$Q = \mathcal{A}/R$. 
We have chosen $L = 10$ and $R = 4$ for both structures and set  
$\gamma = 10$ for the smooth boundary and $\Delta = B(1/2,11/2)L/\pi = 
1.23$ for the sharp boundary in order to have a unique 
$g_H$ in Eq.~\eqref{eq:genDeff}, which means equivalent channels (i.e. 
with the same area). 
The dashed lines in Fig.~\ref{fig:Deff} show the values of   
$D_{\mathrm{eff}}$ of the Sm-channel 
obtained by the LJ-formula~\eqref{eq:chapFive_L_J} 
via a numerical evaluation of the integrals 
for various expressions of the local diffusion coefficient.

The LJ formula cannot be applied to the Sh-channel, 
as $\omega_{\mathrm{Sh}}(x)$ is not differentiable everywhere. 
In this case, another approximation which can be very useful  
is based on the boundary homogenization \citep{Brz2_010,Brz13_06,Brz9_09}.
The method was applied by Berezhkovskii and co--workers \cite{Brz3_07} 
to the calculation of $D_{\mathrm{eff}}$ for the Sh-channel, 
however their result is valid under the restriction 
$2\Delta \ge 2\mathcal{A} + R$ which is not assumed here. 
\begin{figure}
\includegraphics[clip=true,keepaspectratio,width=0.9\columnwidth]
{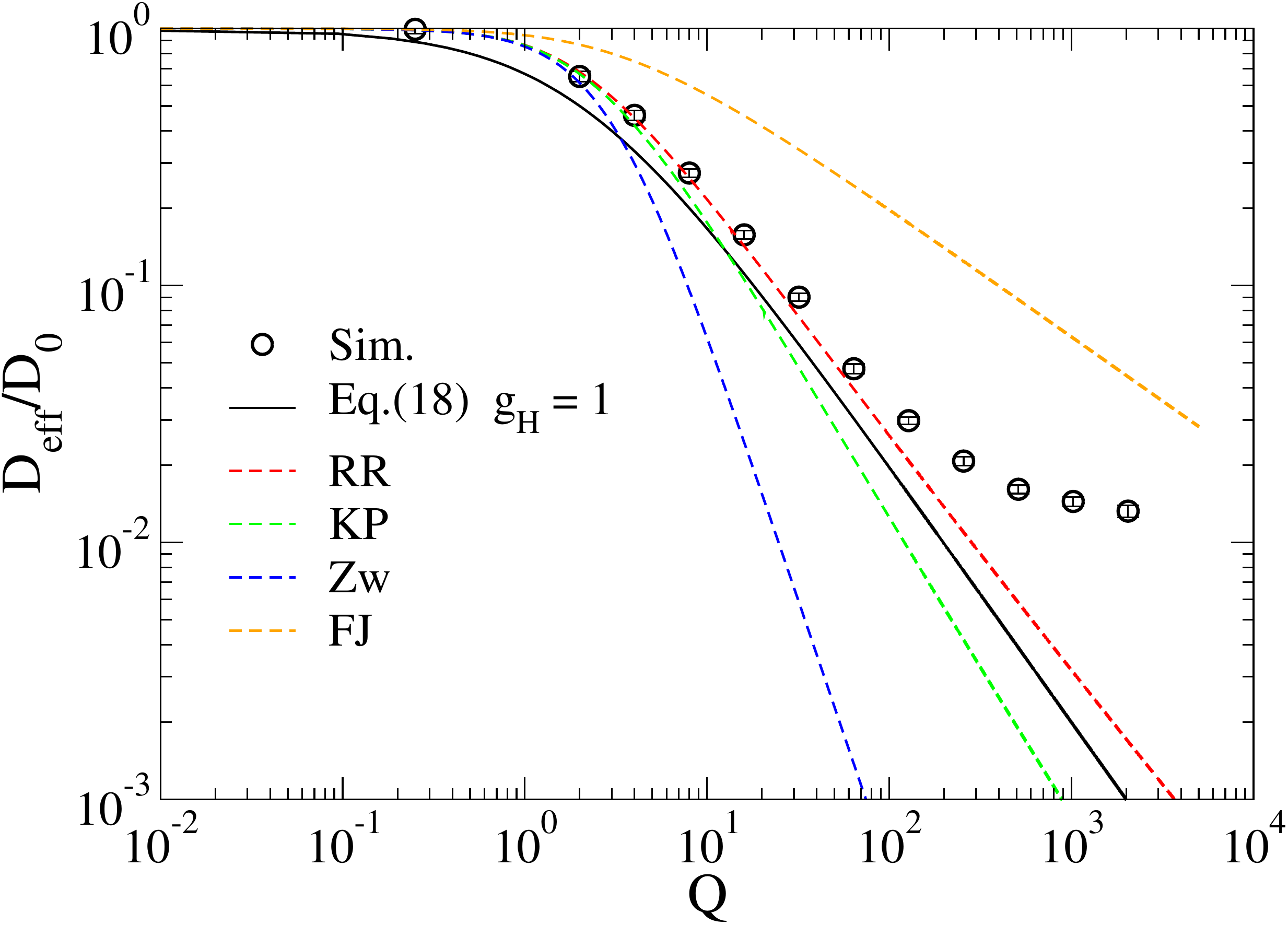}
\caption{\label{fig:Deffgammone}
Comparison between simulated data of $D_{\mathrm{eff}}$ (circles) and 
the theoretical prediction~\eqref{eq:genDeff} (black line) for 
the unbiased diffusion in a Sm--channel, with $\gamma = 1$ and 
the other parameters as in Fig.~\ref{fig:Deff}. 
The percentage error on simulation data is within 4\%.
The accuracy 
of Eq.\eqref{eq:genDeff} worsens at increasing $Q = \mathcal{A}/R$. 
Dashed lines represents the predictions of $D_{\mathrm{eff}}$ 
yielded by the generalized FJ approach involving the same 
$D(x)$ expressions of Fig.~\ref{fig:Deff}.}
\end{figure}
The formula of geometrical areas \eqref{eq:Deff} reveals some
inaccuracy in predicting the $D_{\mathrm{eff}}$ behaviour for the 
Sm-channel with odd $\gamma$, indeed, the case $\gamma=1$ 
reported in Fig.~\ref{fig:Deffgammone} shows that the agreement between 
Eq.~\eqref{eq:genDeff} (solid line) and the simulation data (symbols)
dramatically worsens in the range of large $Q = \mathcal{A}/R$. 

\begin{figure}
\includegraphics[clip=true,keepaspectratio,width=0.8\columnwidth]
{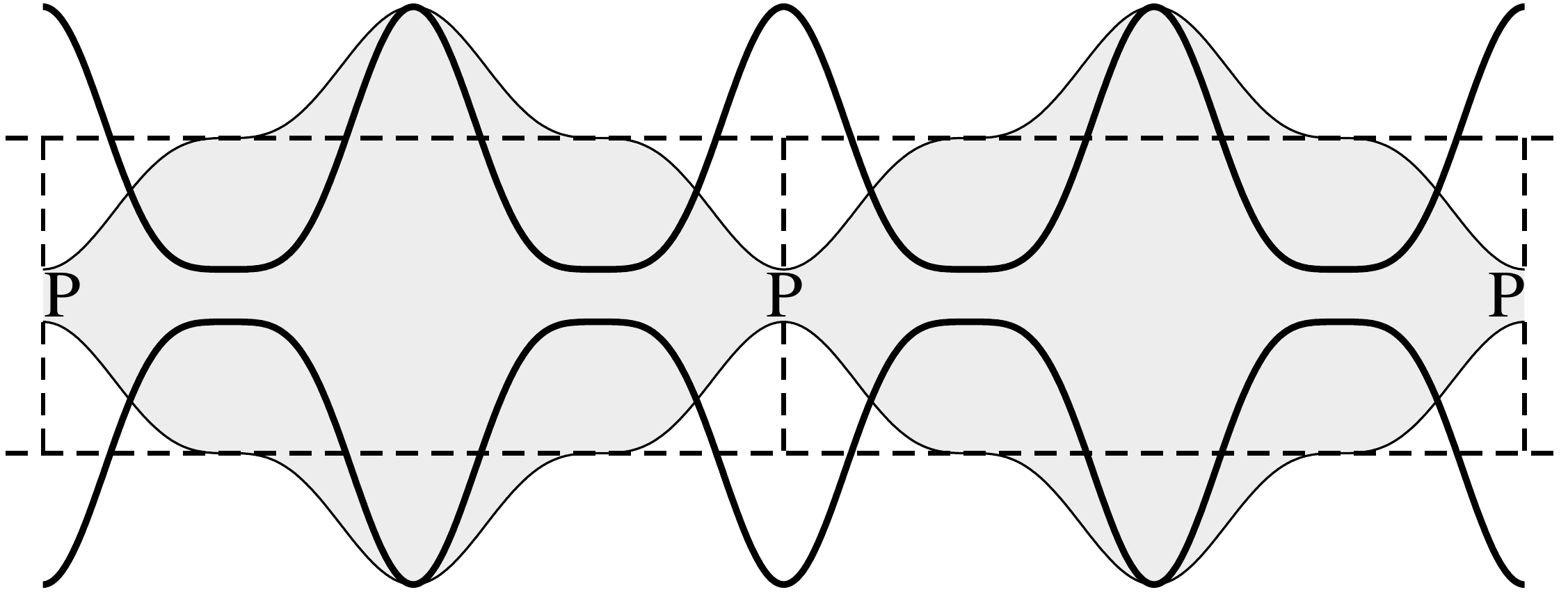}
\caption{\label{fig:septates}
Sketch of a Sm-channel with even (solid) and odd (shaded) $\gamma$. 
Notice the concavity change of the profile when $\gamma$ turns from even
to odd. For $\gamma$ odd, the $H$-region of the channel becomes predominant 
whereas the $S$-domain remains ill-identified as it is 
restricted to point-like regions marked by the letter $P$.
In this case, the structure becomes equivalent to a 
septate channel \cite{Septate2} of section $R+\mathcal{A}$
(dashed lines) for which Eq.~\eqref{eq:Deff} fails.}
\end{figure}

This discrepancy can be justified by observing that 
Eq.~\eqref{eq:genDeff} remains a reliable approximation of 
$D_{\mathrm{eff}}$ as long as the channel partitioning    
in $H$ and $S$ regions is not ambiguous. 
Actually, the $S$ and $H$ distinction is not always physically 
meaningful, just like in the case of the Sm-channel with odd $\gamma$.
The crucial difference between odd and even $\gamma$ is apparent in 
Fig.~\ref{fig:septates}.    
When $\gamma$ is odd, the $S$-domain turns out to be ill defined 
as it shrinks to point-like regions (marked by $P$s in the 
figure). Conversely, the $H$-regions are so large that produce 
a negligible trapping mechanism. The resulting structure is virtually 
equivalent to a ``septate channel'' \cite{Septate2} of section 
$R+\mathcal{A}$,  
namely, an array of compartments connected by holes in the 
separating walls (dashes shape in Fig.~\ref{fig:septates}).  
As it discussed in Ref.~\cite{Mar_5},
the derivation of $D_\mathrm{eff}$ in septates requires a different 
approach from the one leading to Eq.\eqref{eq:genDeff}. 
\begin{figure}
\includegraphics[clip=true,keepaspectratio,width=0.9\columnwidth]
{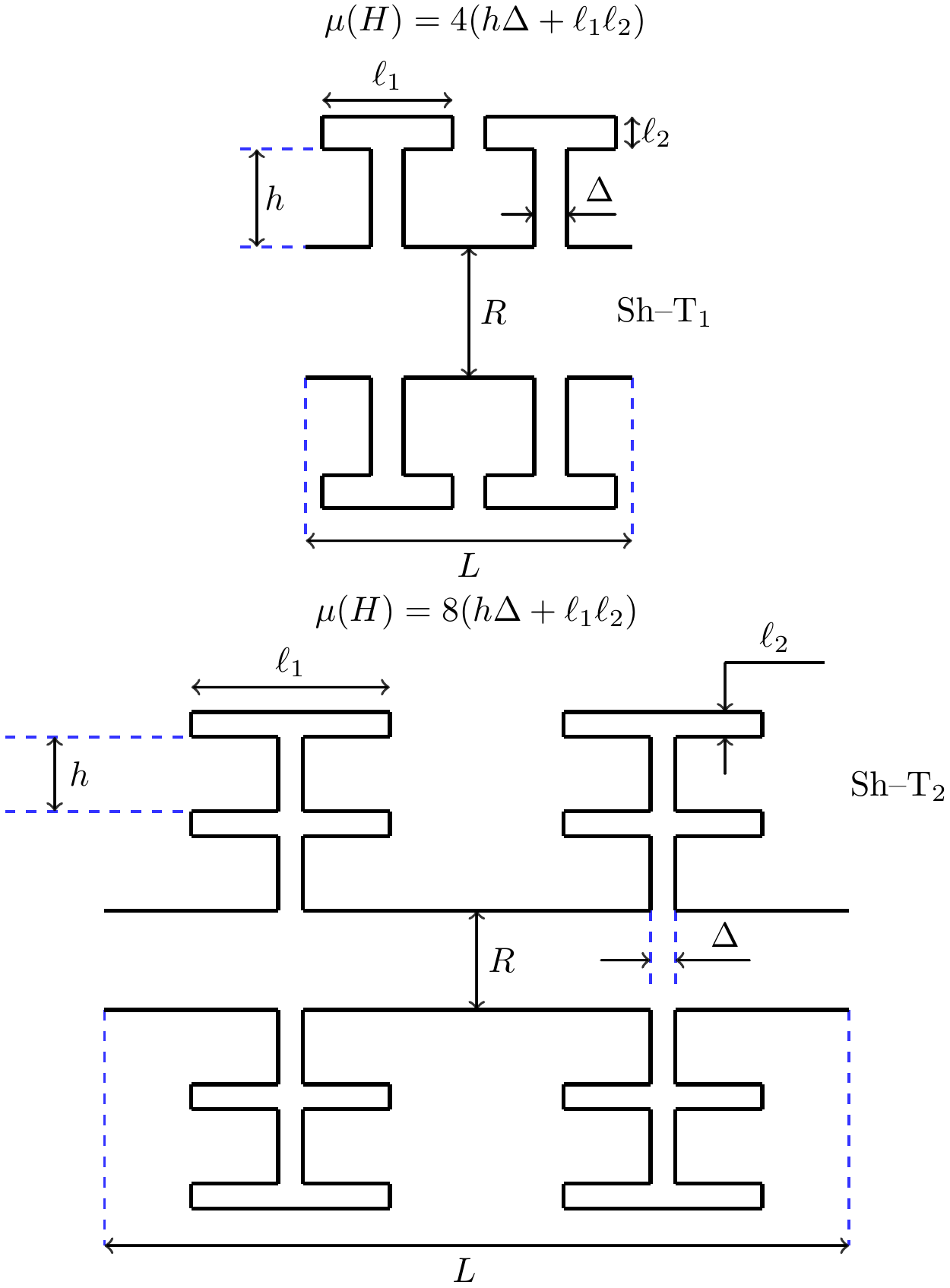} 
\caption{Examples of sharp channels Sh-T$_1$ and Sh-T$_2$ with a 
boundary profile $\omega(x)$ which is a multivalued function of 
the longitudinal position; in that case the generalized FJ 
approximation cannot be used. 
In both structures, the parameters, $h,\Delta,\ell_1,\ell_2$ are tuned   
such that $\mu(H)$ coincides with the Hump area 
of the boundaries used to generate Fig.~\ref{fig:Deff}.}
\label{fig:otherstruct}
\end{figure}

\begin{figure}
\includegraphics[clip=true,keepaspectratio,width=0.9\columnwidth]
{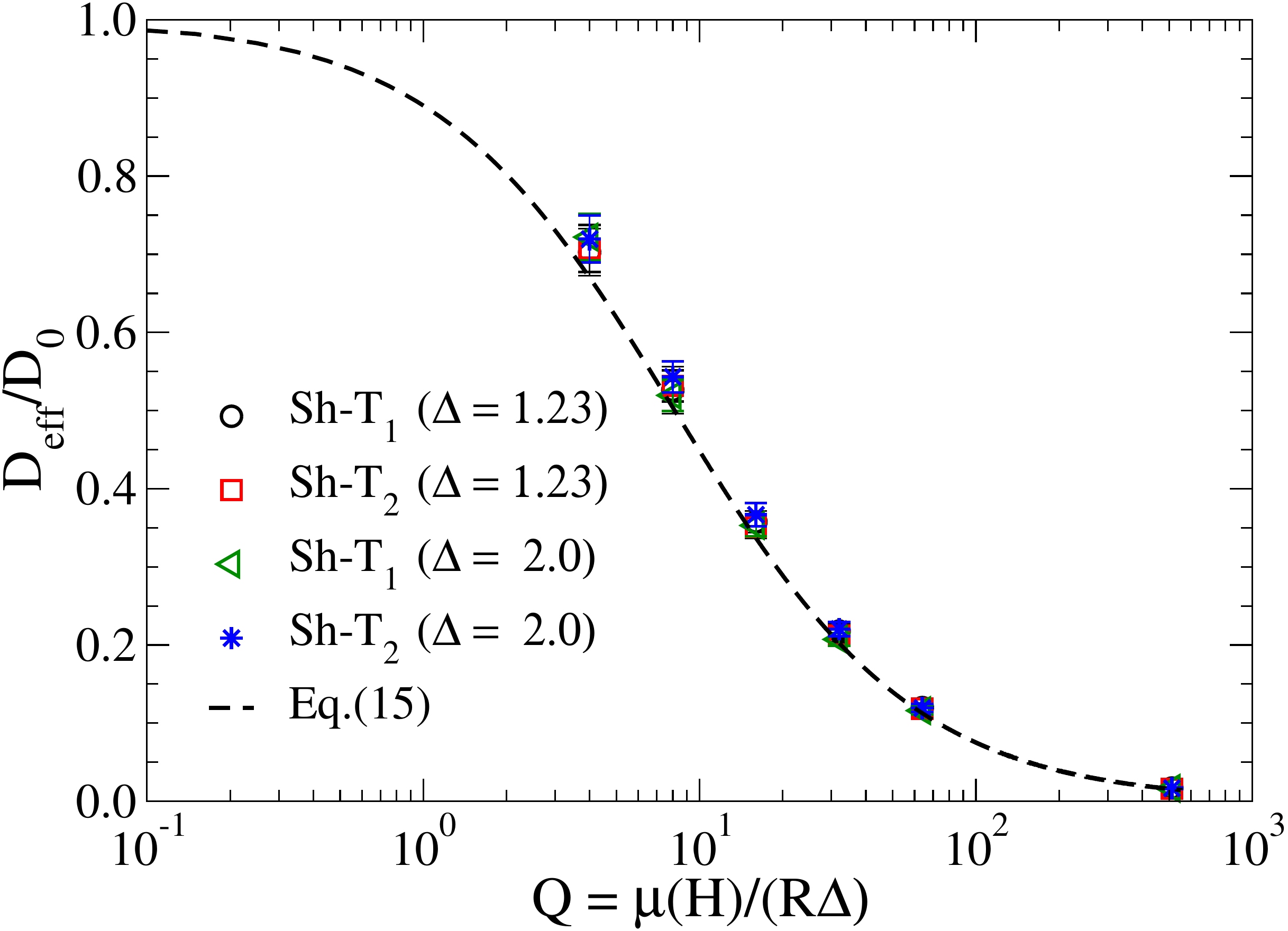}
\caption{\label{fig:Defficaci}
Plot of $D_{\mathrm{eff}}/D_0$, theory (dashed line) and 
simulation data (symbols) vs $Q = \mu(H)/(R\Delta)$ referred to 
the structures Sh-T$_1$ and Sh-T$_1$ drawn in Fig.~\ref{fig:otherstruct}. 
We considered channels with two $\Delta$ values and tuned the other 
parameters ($h,\ell_1,\ell_2$) to maintain the same Hump area 
in both structures. For this reason different data sets collapse 
onto each other.  
The agreement between theory and simulation results 
is convincing despite the geometric complexity of the channels. 
The percentage error on data is about 4\% leaving the bars well 
below the symbol size.}
\end{figure}

Apart from the peculiarity of the septates, Eq.\eqref{eq:Deff} is 
rather general as it works well in a variety of geometries. 
For instance, it is able to predict the $D_{\mathrm{eff}}$ associated 
with the complex channels Sh-T$_1$ and Sh-T$_2$ sketched in 
Fig.~\ref{fig:otherstruct}. 
This can be appreciated from the results in 
Fig.~\ref{fig:Defficaci} showing a   
comparison of $D_{\mathrm{eff}}$ from Brownian simulations 
in Sh-T$_1$ and Sh-T$_2$ (symbols) to its theoretical prediction 
based on Eq.\eqref{eq:Deff} (dashed line). The different sets of 
data refer to structures with different profiles but the same hump 
area $\mu(H)$.  
The perfect collapse of the different data  
onto the same theoretical curve confirms the great accuracy of 
Eq.\eqref{eq:genDeff}. 

This example clearly indicates that, when the  
$S$-$H$ partition of a channel is not ambiguous 
and the $H$-regions perform a sufficient trapping action on 
the particles, the area's formula~\eqref{eq:genDeff} is able to  
catch the right dependence of $D_{\mathrm{eff}}$ on the channel 
geometry, despite the complexity of the section.  

\section{Pre--asymptotic diffusion
\label{sec:preasy_new}}
Generally, the study of pre--asymptotic or transient regimes 
is important to gain information about a possible trapping mechanism 
and memory effects in a dynamical processes \cite{artale1997dispersion,ammenti2009}.
In the specific case of corrugated channel, the temporary 
trapping of particles within the $H$ regions can  
induce a transient behaviour
which generally depends on the system preparation and 
it could be characterized by nonlinear growth of the mean 
square displacement \cite{Tanner78,Brz10_07}.  
There exists a crossover time $\tau_{\parallel}$  separating the 
pre--asymptotic and the asymptotic motion  
that is affected by the initial particle 
distribution in the channel, 
$\tau_{\parallel} = \tau_{\parallel}(\mathbf{r}_{0})$,  
$\mathbf{r}_0$ being a shorthand notation
for the initial position of all the particles.
We considered three types of initial conditions in the channel 
of Fig.~\ref{fig:Chann}b: 
i) a uniform distribution 
of particles in the $S$-region of a single unitary cell, 
$\mathbf{r}_{0}\in S$, ii) the particles uniformly distributed 
in one period of the $H$-region, 
$\mathbf{r}_{0}\in H$, and finally iii) a uniform distribution over 
the whole unit cell $\mathbf{r}_{0}\in S\cup H$.
\begin{figure}
\centering
\includegraphics[clip=true,keepaspectratio,width=0.9\columnwidth]
{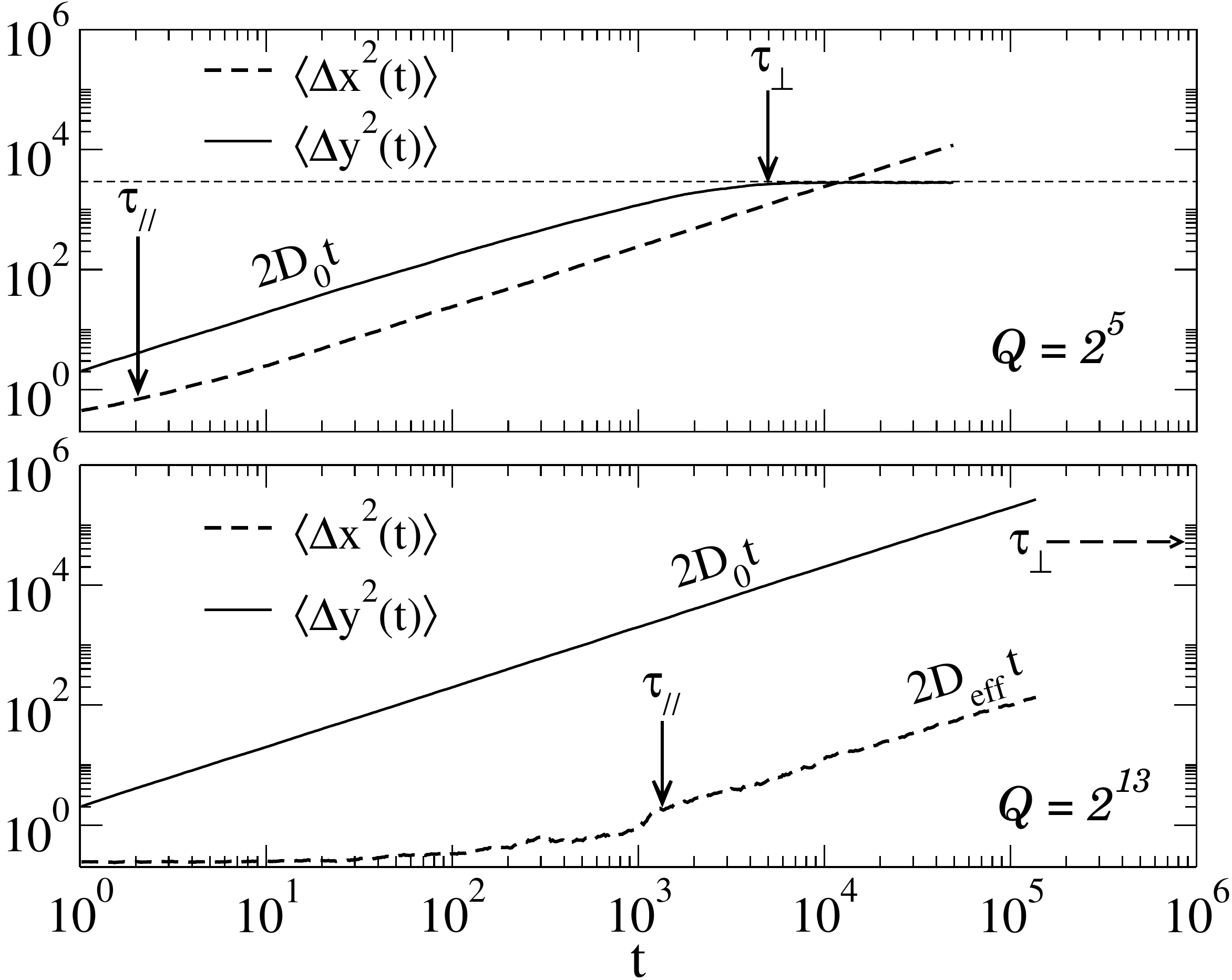}
\caption{\label{fig:PreAsi_y} Comparison 
between the relaxation of longitudinal $\langle \Delta x(t)^{2}\rangle$ 
and transversal $\langle \Delta y^2(t)\rangle$ diffusion 
for the same channel of Fig.~\ref{fig:PreAsi} with two  
different values of $Q = 2^5$ and $Q = 2^{13}$. 
The system is prepared uniform over $S\cup H$, but 
the large $Q$ values prevents the fast lateral homogenization.
In the case $Q = 2^5$, the homogenization takes place at times  
$\tau_{\parallel} < \tau_{\perp}$, while for 
$Q = 2^{13}$ the homogenization becomes particularly slow    
($\tau_{\perp} \simeq 3\time 10^9$) and it delays the onset 
of the standard diffusion along the $x$-axis, $\tau_{\perp}\simeq 10^3$. 
Moreover as long as $t < \tau_{\perp}$, the lateral diffusion satisfies  
$\langle\Delta y^2(t)\rangle = 2D_0 t$, 
supporting the derivation of Eq.\eqref{eq:Pt} and Eq.\eqref{eq:MSDt}.
The horizontal dashed line marks the value 
$2 D_0\tau_{\perp}(S\cup H) = R^2(1+Q)^2/6$ obtained from the last line 
of Eq.\eqref{eq:Tperp}.}
\end{figure}
The ``local equilibrium'' assumption \citep{Zw1_92,Mar_3} requires 
that the typical relaxation time 
$\tau_{\perp}$ after which the particle
distribution becomes uniform in each cross section is 
much smaller than $\tau_{\parallel} \gg \tau_{\perp}$ 
where $\tau_{\perp}$ is estimated assuming a flat final 
$y$ distribution on each section $[-\omega(x),\omega(x)]$, 
$$
\tau_{\perp} = \frac{\langle[y(\infty) - y(0)]^2\rangle}{2D_0} =  
\frac{\sup_x \{\omega^2(x)\}}{6D_0} + \frac{\langle y^2(0)\rangle}{2D_0},
$$
the angular brackets indicate the double average over the final and
initial distributions.
In the first term,   
$\sup_x \{\omega^2(x)\} = (R+\mathcal{A})^2/4$ for both Sm and Sh channels, 
the value $\langle y^2(0)\rangle$ depends on the starting particle 
distribution. 
In particular, for the distributions i)--iii) just mentioned we have,
\begin{equation}
\tau_{\perp}(\mathbf{r}_0) = \tau_R
\begin{cases}
(1+Q)^2 + 1 \qquad \mathbf{r}_0 \in S 
\vspace{0.5em}\\ 
(1+Q)^2 +\dfrac{(1+Q)^3-1}{Q} \quad \mathbf{r}_0 \in H
\vspace{0.5em}\\
2(1+Q)^2 \qquad \mathbf{r}_0 \in S\cup H. 
\end{cases}
\label{eq:Tperp}
\end{equation}
with $\tau_R = R^2/(12D_0)$.
Taking $Q$ sufficiently large, $\tau_{\perp}$ can be made
so arbitrarily long that the transversal homogenization 
becomes considerably slow and the assumption of a sharp scale 
separation $\tau_{\perp} \ll \tau_{\parallel}$ certainly fails.
This very slow transversal homogenization induces a delay on the onset of 
the standard longitudinal diffusion that can be observed 
in Fig.~\ref{fig:PreAsi_y}, where we plot the relaxation of 
longitudinal and transversal MSD in a sharp channel for a 
moderate $Q = 2^7$ and large $Q = 2^{15}$. 
In both cases, the vertical homogenization 
established later than the occurrence of the normal diffusion along 
the channel axis.
However in the second case, the onset of a longitudinal standard regime 
is strongly retarded by the lack of lateral homogenization. 
 
\begin{figure}
\includegraphics[clip=true,keepaspectratio,width=0.9\columnwidth]
{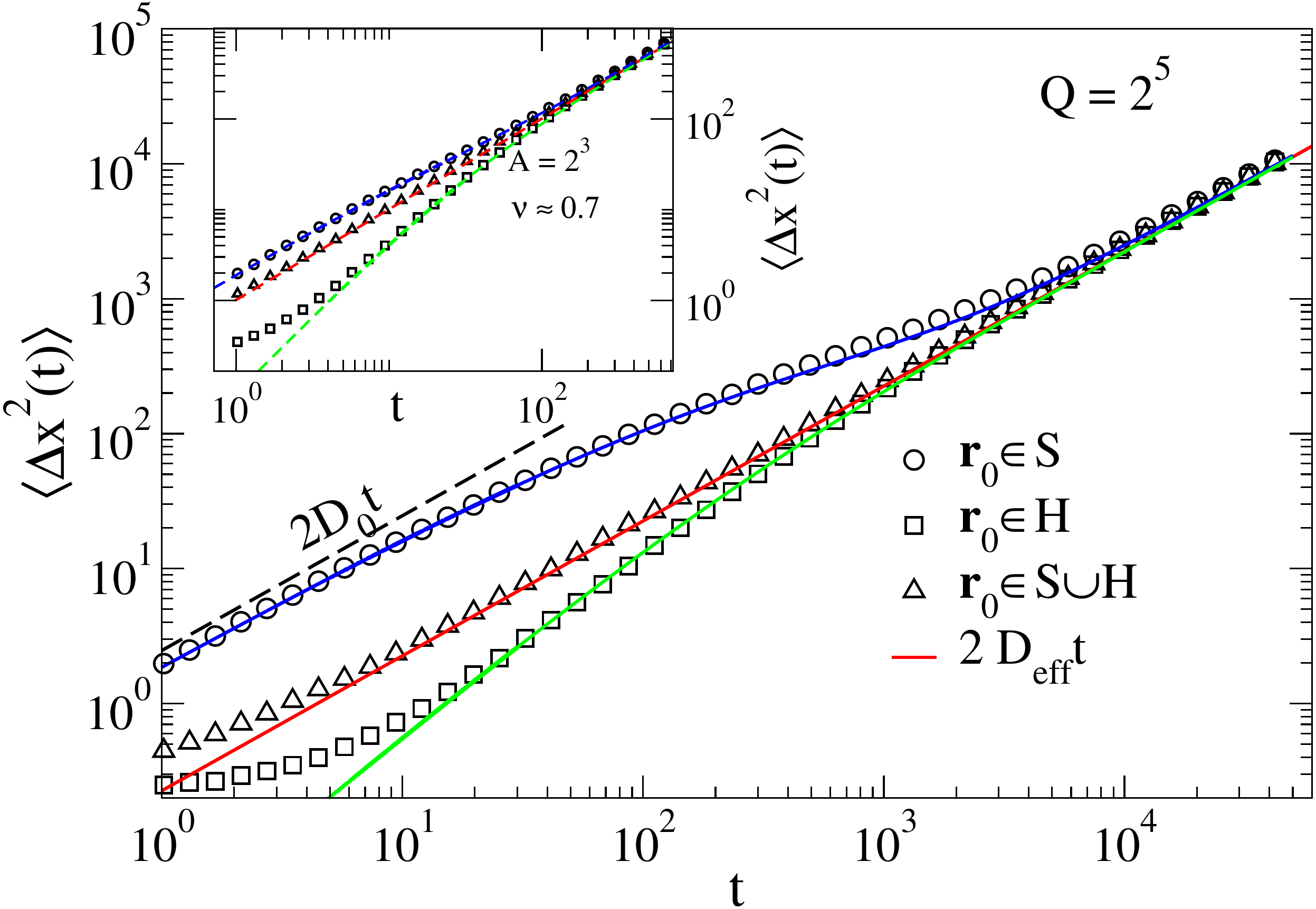}
\caption{\label{fig:PreAsi} Log--log plot of 
$\langle\Delta x^2(t)\rangle$ in a Sh-channel with $Q = 2^{5}$     
as a function of the time for the system starting from three 
initial conditions: $\mathbf{r}_{0}\in H$ (squares),
$\mathbf{r}_{0}\in H\cup S$ (triangles), 
$\mathbf{r}_{0}\in S$ (circles). 
Symbols refer to the simulation data whereas the dashed lines represent 
their fit by Eq.~\eqref{eq:MSDt} upon tuning the free parameter $\tau_{0}$.
Parameters $R, L,\Delta$ are as in Fig.~\ref{fig:Deff}.
The inset shows the same results for $Q =2^{2}$.} 
\end{figure}

Therefore, we shall focus on the pre--asymptotic diffusion 
in channels with large enough $Q$, where, more likely, the local equilibrium 
condition is violated.  
Figure~\ref{fig:PreAsi} shows the log--log plot of the longitudinal MSD 
for the three initial distributions i)--iii) defined above in 
the Sh-channel~\eqref{eq:Rec_Chann_w} with  $Q = 2^{5}$, 
the inset reports the same data for $Q = 2^{2}$. 
The plot emphasizes the transient nonlinear behaviour of MSD and 
the dependence of the dynamics on the initial condition. 
The different short-time behaviour of the three sets of data 
can be explained as follows. The particles initialized with    
$\mathbf{r}_{0}\in S$  
perform a stage of free diffusion (coefficient $D_0=1$) until they are 
trapped into the Humps, accordingly their MSD initially grows like 
$\langle \Delta x^2(t)\rangle \simeq 2D_0 t$ (circles).
The opposite case corresponds to all the particles initialized in 
$\mathbf{r}_{0}\in H$. 
Their MSD remains stacked to a small value (squares) due to the trapping 
of Humps until enough particles begin to escape.
Obviously, the MSD of particles uniformly initialized in 
$\mathbf{r}_{0}\in H\cup S$ exhibits a short-time behaviour 
that is intermediate between the two opposite cases. 

The global behaviour of $\langle \Delta x^2(t)\rangle$
as a function of the elapsed time can be qualitatively understood
by the following argument. 
The growth $\langle \Delta x^2(t)\rangle$ 
can be related to the occupation probability $P_S(t)$ 
of the $S$ region at time $t$ \cite{Brz10_07} 
\begin{equation}
\langle \Delta x(t)^{2}\rangle = 2 D_{0} \int_{0}^{t}\! du\;P_{S}(u)
\label{eq:Dt}
\end{equation}
In the Appendix~\ref{app:A}, we show how Eq.~\eqref{eq:Dt}, 
which generalizes Eq.~\eqref{eq:Deff} to transient diffusion,
can be justified by exploiting some intuitive similarities between the 
diffusion in corrugated 
channels and the unbiased random walks on a comb lattice with finite 
sidebranches. 
Equation~\eqref{eq:Dt} is, of course, meaningful only if 
the $S$ and $H$ regions can be unambiguously identified. If so, 
we can define a coarse--grained description of the process in 
terms of a two-state kinetic model, where the states $S$ and $H$ 
have probabilities $P_{S}(t)$ and $P_{H}(t)=1-P_{S}(t)$, respectively. 
Let $k_{\mathrm{S}}(t)$ and $k_{\mathrm{H}}(t)$ be the transition rates 
from $S$ to $H$ and from $H$ to $S$ respectively,  
then the rate equation for $P_{S}(t)$ reads
\begin{equation}
\frac{d P_{S}(t)}{dt} = -k_{\mathrm{S}}(t) P_{S}(t) + 
                         k_{\mathrm{H}}(t) [1-P_{S}(t)]. 
\label{eq:coarsePt}
\end{equation}
We considered the limiting case of large $\mathcal{A}$,  
in which the strong trapping mechanism delays the longitudinal 
equilibration (homogenization), see Fig.~\ref{fig:PreAsi_y}.
Accordingly, for $t\ll\tau_{\perp}$ the diffusion in the $y$ direction
is still an ongoing process, $\langle\Delta y^{2}(t) \rangle \sim t$. 
In a first approximation, we can argue that  
$k_S(t)$ and $k_H(t)$ are proportional to the typical 
spreading velocity of the transversal diffusion, thus,   
$$
k_{\mathrm{S}}(t) \sim a/\sqrt{t}, \qquad k_{\mathrm{H}}(t) \sim b/\sqrt{t};  
$$
where $a$ and $b$ are dimensional constants to be determined by 
phenomenological considerations. 
Substituting the rates into Eq.~\eqref{eq:coarsePt}, we get  
the solution 
$$
P_S(t) = P_{S}(0)\mathrm{e}^{-2(a+b)\sqrt{t}} + 
\frac{b}{a+b} \left(1 - \mathrm{e}^{-2(a+b)\sqrt{t}}\;\right).
$$
Now the parameters $a$ and $b$ can be expressed in terms of meaningful
physical quantities by setting: 
$2(a+b) = 1/\sqrt{\tau_0}$,  
where $\tau_{0}$ is a free time scale to be adjusted, and 
$b/(a+b) = D_{\mathrm{eff}}/D_{0}$, 
which comes from the equilibrium condition 
$P_S(t) \to P_S^{eq}$. 
After simple algebra we obtain   
$a = (1 - D_{\mathrm{eff}}/D_{0})/\sqrt{4\tau_0}$ and 
$b = D_{\mathrm{eff}}/(D_{0}\sqrt{4\tau_0})$.   

Thus, the final expression for $P_S(t)$ reads   
\begin{equation}
P_{S}(t) =    
P_{S}(0) \mathrm{e}^{-\sqrt{t/\tau_0}} + 
    \frac{D_{\mathrm{eff}}}{D_0} \left(1 - \mathrm{e}^{-\sqrt{t/\tau_0}}\;\right)
\label{eq:Pt}
\end{equation}
and according to Eq.\eqref{eq:Dt}, we derive the MSD 
\begin{equation}
\langle \Delta x(t)^{2}\rangle = 2 D_{\mathrm{eff}} t +   
C\left\{1 - \mathrm{e}^{-\sqrt{t/\tau_0}}\left(\sqrt{t/\tau_0} + 1\right)\right\}, 
\label{eq:MSDt}
\end{equation}
with $C = 4D_0\tau_0 [P_S(0) - D_{\mathrm{eff}}/D_0]$.
The phenomenological expression \eqref{eq:MSDt} can be used to fit  
the MSD data of Fig.~\ref{fig:PreAsi} by adjusting the $\tau_{0}$ 
value. 
The blue dashed line represents Eq.\eqref{eq:MSDt} 
with the initial condition $P_{S}(0)=1$, corresponding to all the 
particles in the $S$ region. Since $C$ is positive, the second term 
in Eq.\eqref{eq:MSDt} contributes positively to the MSD and 
the convergence to the asymptotic behaviour $2 D_{\mathrm{eff}} t$ is from 
above.      
When the particles start in the $H$ region, $P_{S}(0)=0$,  
Eq.~\eqref{eq:MSDt} corresponds to the dashed green line, in this case 
$C<0$ and the convergence to the slope $2 D_{\mathrm{eff}} t$ is from below.
Finally, if the particles are initialized with a uniform distribution 
so that $P_{S}(0) = P_{S}^{eq} = D_{\mathrm{eff}}/D_0$,      
the constant $C$ in Eq.~\eqref{eq:MSDt} vanishes and the 
system achieves soon the expected standard diffusive behaviour 
(red dashed line). 
The agreement between simulation data and the fitting curves is satisfactory 
considering that Eq.~\eqref{eq:MSDt} has only one free parameter  
and that its derivation is based just on reasonable assumptions. 

The regime with $\mathcal{A}$ comparable with all the other 
geometrical sizes can be captured by a generalization of 
Eq.~\eqref{eq:Pt} where the square-root dependence 
$\sqrt{t/\tau_0}$ is replaced by the 
power law $(t/\tau_0)^{\nu}$, with $\nu$ being an additional fitting parameter.
The inset of Fig.~\ref{fig:PreAsi} shows the example with 
$\mathcal{A} = 2^{3}$, 
for which we found $\tau_{0}\approx 18$ and $\nu\approx 0.7$.

\section{Einstein's Fluctuation Dissipation Relation.}
\label{sec:FDR}
We discuss the transport problem 
in the presence of small external longitudinal field 
$\mathbf{F}=f\hat{\mathbf{x}}$, i.e.,  $f L/(2 k_{\mathrm{B}}T)\ll 1$. 
In particular, we are interested in possible generalization of the 
Einstein's (FDR) to regimes where the 
approach to the ``transversal equilibrium'' is so slow that a robust 
transient behaviour of non standard longitudinal diffusion can be 
observed.

In the linear response regime, the asymptotic 
mean drift induced by the field $f$ is  
$$
\langle \Delta x(t) \rangle_{f} = \mu_{\mathrm{eff}} f t, 
$$
where $\mu_{\mathrm{eff}}$ is the effective mobility, 
and since the asymptotic MSD for $f=0$ is given by
$$
\langle \Delta x^2(t) \rangle_{0} = 2D_{\mathrm{eff}} t,
$$ 
the ratio 
\begin{equation}
\frac{\langle\Delta x(t)\rangle_{f}}{\langle \Delta x^2(t)\rangle_{0}}=
\frac{\mu_{\mathrm{eff}}}{D_{\mathrm{eff}}} \frac{f}{2}  
= \mbox{const}.
\label{eq:finalFDR}
\end{equation}
remains constant in time, assuming Einstein's FDR
$D_{\mathrm{eff}} = k_B T \mu_{\mathrm{eff}}$, the const is fixed to 
$f/(2 k_B T)$. Here we work in units such that $k_BT = 1$.  

\begin{figure}
\includegraphics[clip=true,keepaspectratio,width=0.9\columnwidth]
{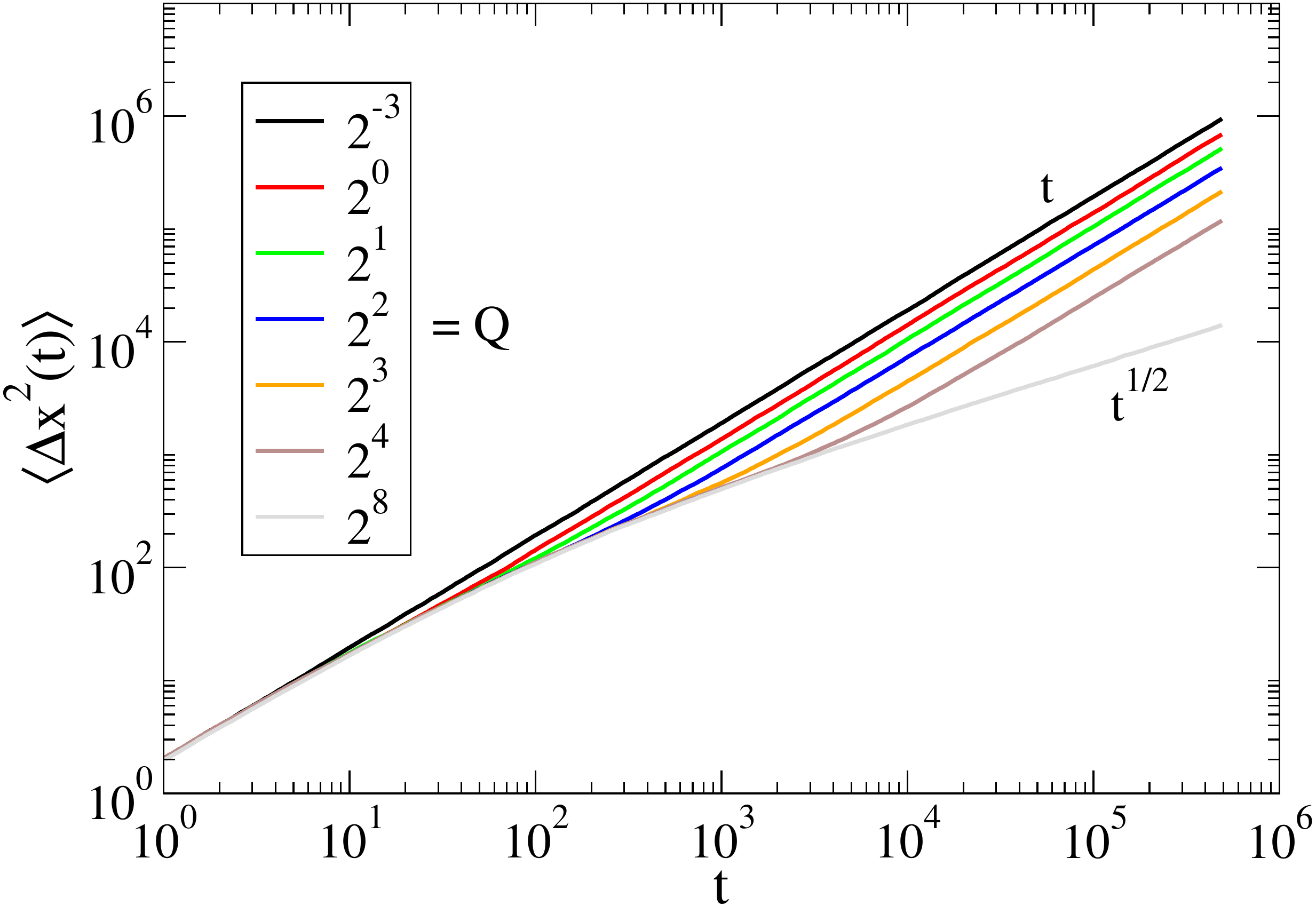}
\caption{\label{fig:Xquadri} Log--log plot of the MSD 
versus the elapsed time for the channel sketched in 
Fig.~\ref{fig:Chann}(b) with increasing values of $\mathcal{A}$.
For $\mathcal{A}/R = 2^8$, the diffusion is anomalous
$\langle \Delta x^2(t)\rangle_0 \sim t^{1/2}$ with the same 
exponent of the random walk on a comb, $2\nu=1/2$, supporting   
the view that the structure in Fig.~\ref{fig:Chann}(b) virtually behaves as a 
comb at enough large $\mathcal{A}$.} 
\end{figure}

\begin{figure}
\includegraphics[clip=true,keepaspectratio,width=0.9\columnwidth]
{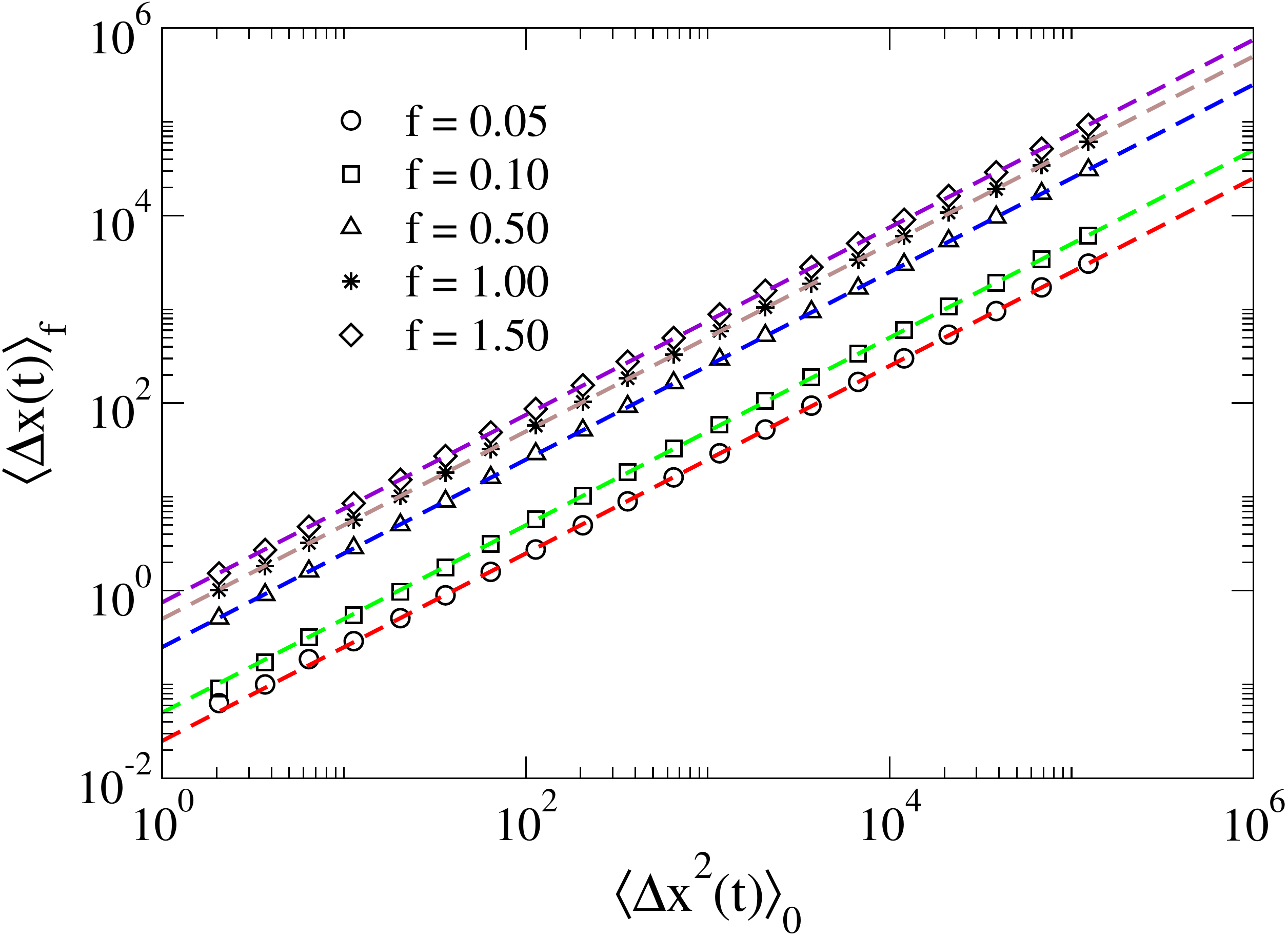}
\caption{Check of the FDR for the diffusion 
in Sh-channels-\eqref{eq:Rec_Chann_w}. Log--log plot of
$\langle\Delta x(t)\rangle_{f}$ vs $\langle \Delta x^2(t)\rangle_{0}$, 
symbols are the simulation results, and dashed lines represent the 
theoretical expectation~\eqref{eq:finalFDR}, which is verified 
regardless of the $\mathcal{A}$ value and turns out to be independent of 
the initial particle distribution within the channel. 
The FDR is exactly satisfied also in smooth channels (not shown).}
\label{fig:Response} 
\end{figure}

Previous studies \cite{Hanggi1_07,Mar_5,Mar_7,Berez_2011} mainly focused 
on the validity of FDR in the asymptotic regime and worked out  
analytical expressions for the linear and nonlinear mobility 
$\mu_{\mathrm{eff}}$, here we test Eq.\eqref{eq:finalFDR} when the 
diffusion is not yet asymptotic. 
This analysis can be guided by the structural similarity
that channels with $\mathcal{A}/R \gg 1$ share with the comb lattice
with long but finite sidebranches (Fig.~\ref{fig:comb} of 
the Appendix~\ref{app:A}):
the backbone and teeth of the comb play the role of
the $S$ and $H$ regions of the channel.   
This similarity reflects also on the transport properties, indeed 
Fig.~\ref{fig:Xquadri} shows 
that in the limit $Q = \mathcal{A}/R \gg 1$ the diffusion along the 
channel becomes anomalous $\langle \Delta x^2(t)\rangle_0 \sim t^{2\nu}$ 
with the same exponent $\nu = 1/4$ of the random walk on a comb 
\cite{Havlin87,Weiss86,Forte013}. 
This scenario is consistent with the behavior 
$D_{\mathrm{eff}} \sim 1/\mathcal{A}$ at large $\mathcal{A}$ [see  
Eqs.\eqref{eq:sinDeff} and \eqref{eq:rectDeff}], which is    
a physical consequence of the trapping action exerted on particles 
by very long humps.
The Appendix~\ref{app:A} shows that on a comb lattice such a regime, 
though anomalous, does satisfy the FDR at any time due
to an accidental but exact cancellation in the ratio~\eqref{eq:finalFDR}. 
We stress that in the comb lattice, Eq.\eqref{eq:finalFDR} is  
exact not only at any times but also for any initial particle distribution. 
In analogy, as long as the random walk on the comb constitutes a good 
representation of the continuum process,   
we expect FDR to maintain a certain validity for the 
diffusion in channels even in non-asymptotic regimes and 
regardless of the initial particle distribution, provided it is 
generic enough. 
To verify this, we performed numerical simulations of diffusion 
driven by different external fields and in channels with increasing 
$\mathcal{A}/R$.
The plot of $\langle \Delta x(t)\rangle_f$ versus 
$\langle \Delta x^2(t)\rangle_0$ for different values of 
$Q = \mathcal{A}/R$ is shown in Fig~\ref{fig:Response}, where 
symbols are the numerical data and dashed straight lines represent the 
expected behaviour as predicted by taking the logarithm of both members of 
Eq.\eqref{eq:fdt_app}. 
The perfect alignment of data along the 
lines confirms the exact proportionality between the MSD and 
the mean drift according to Eq.\eqref{eq:finalFDR}.

We note that in systems governed by a one-dimensional 
Fokker-Planck equation, the Einstein's relation holds exactly at 
any time if the initial condition coincides with the invariant
probability distribution \cite{BettoloRep}. The same applies  
also to the fractional Fokker-Planck equation \cite{MetzlerPRL99}
describing anomalous diffusion.    

In conclusion,   
although both $\langle\Delta x(t)\rangle_{f}$ and
$\langle \Delta x^2(t)\rangle_{0}$ depend on the channel shape 
and on the initial system preparation, 
their ratio remains constant in time in any explored regime, 
generalizing ``{\em de facto}'' the Einstein relation to the pre-asymptotic
dynamics. 

\section{Conclusions}
\label{sec:conclusion}
This paper analyzed the diffusion in channels with nonuniform sections
in conditions well beyond the range of applicability of the FJ 
approximation, which properly works only for small deviations 
from the homogeneous channel. 
We performed Brownian dynamics simulations in different geometries 
considering even channels with steep variations of the 
boundaries and computed numerically the effective diffusion coefficient 
$D_{\mathrm{eff}}$ from the 
asymptotic behavior of the mean square displacement (MSD).   
We compared the numerical results to the predictions from 
the various corrections to the FJ theory, suggested in the literature, 
that make use of a local diffusion coefficient within the 
Lifson--Jackson formula. 
We then focused on channel profiles which are not differentiable 
(sharp channels) for which the FJ approximation and its corrections 
turns out to be of difficult application. 
In these cases, there exists a semi-heuristic approach
based on the large-time limit of formula \eqref{eq:Dt} which allows 
$D_{\mathrm{eff}}$ to be expressed in terms of the geometrical 
weight of region $H$, where particles contribute to the 
longitudinal diffusion and dead-end region $S$, where particles 
are temporarily trapped. 
We tested numerically the applicability of Eq.~\eqref{eq:Dt} to 
different channel geometries and we found that it works fine only 
when the partition of the channel into $S$ and $H$ regions is 
unambiguous. 
In other words, the trapping mechanism of the dead-ends should be 
strong enough to introduce a scale separation 
between moving and trapped dynamics. When 
this scale separation does not occur the theoretical prediction deviates
from the numerical data. 

In addition, we characterized the transient regime of the diffusive 
transport by measuring nonlinear behaviour of the MSD 
before it attains the linear growth.  
These transients for specific channel profiles 
can be made arbitrarily long and robust by acting on the channel 
geometry.
Also in this case we analyzed the data according to 
formula \eqref{eq:Dt} obtaining a satisfactory description of the
MSD growth in terms of the solution of phenomenological 
rate equation for the probability to occupy the $S$-region.  
We proved that Einstein's relation can be remarkably 
established between the mean drift of 
the biased diffusion in a small field along the channel 
axis and the MSD of the unbiased diffusion,  
in both asymptotic and pre-asymptotic regimes.

\bibliographystyle{unsrt}
\bibliography{Biblio}

\appendix 
\section{}\label{app:A} 
\begin{figure}
\includegraphics[clip=true,keepaspectratio,width=0.9\columnwidth]
{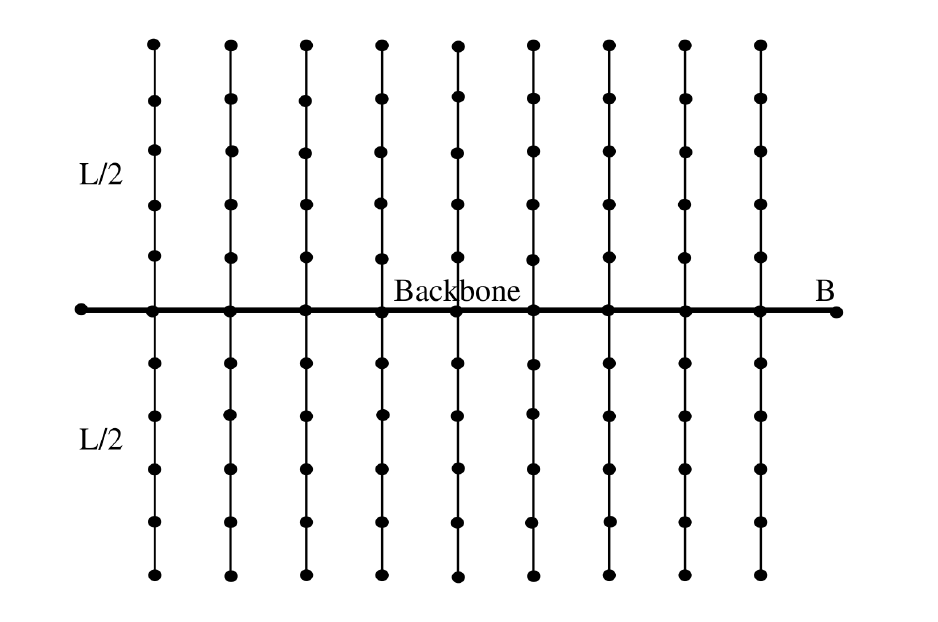}
\caption{Sketch of the regular comb lattice used for its analogy 
to channels with narrow and long hump regions. 
The Backbone plays the role of the channel $S$-region 
while the teeth act as $H$-regions.}
\label{fig:comb}
\end{figure}

This appendix shows the derivation of formula~\eqref{eq:Dt} and
the response by exploiting the analogy between 
the diffusion in a channel geometry with very large $\mathcal{A}$ 
and the random walk on the comb lattice with finite length sidebranches 
(Fig.~\ref{fig:comb}). 

The comb lattice is composed of periodic arrangements of transversal 
teeth of length $L$, attached to the backbone of the structure, which 
represents the transport direction as shown in Fig.~\ref{fig:comb}. 
For simplicity, we consider here the case of unitary spaced teeth.
The sidebranches in comb lattice mimics,  
to some extent, the trapping mechanism of ``Hump'' regions ($H$) of 
the channel.  The backbone corresponds to the region ``$S$'' 
where diffusion is not hindered and the drift is efficient.
Denoting by $\mathbf{r}_{t}=\{x(t),y(t)\}$ the position vector of a 
given particle, the total displacement up to time $t$ along the backbone 
is given by
\begin{equation}
x(t) - x(0) = \sum_{j=1}^{t}\xi_{j}\delta_{y_{j},0},
\label{eq:chapFour_combDIS}
\end{equation}
where the random variable $\xi_{j}$ takes values in $\{1,0,-1\}$, 
respectively with probability $\{1/4,1/2,1/4\}$ \cite{Forte013}. 
The average square displacement is
\eqref{eq:chapFour_combDIS} 
\begin{eqnarray}
\langle[x(t)-x(0)]^{2}\rangle_{0} = 
\sum_{j=1}^{t}\langle(\xi_{j}\delta_{y_{j},0})^{2}\rangle + 
\nonumber
\\
2\sum_{j=1}^{t}\sum_{i>j}^{t}\langle\xi_{j}\xi_{i}\delta_{y_{j},0}
\delta_{y_{i},0}\rangle\;.
\end{eqnarray}
All the terms  $\langle(\xi_{j}\delta_{y_{j},0})^{2}\rangle$ vanish
if $y_{j}\ne 0$, i.e., if a walker is out of the backbone, otherwise
$\langle (\xi_{j}\delta_{y_{j},0})^{2}\rangle = 
\langle\xi^{2}_{j}\rangle=1/2$. 
On the other hand 
$\langle\xi_{j}\xi_{i}\delta_{y_{j},0}\delta_{y_{i},0}\rangle = 0$ if
$j\ne i$. 
We can write
\begin{equation}
\langle[x(t)-x(0)]^{2}\rangle_{0} =  2 tF_{\mathrm{B}}(t),
\label{eq:msd_backbone}
\end{equation} 
with $F_{\mathrm{B}}(t)$ the mean percentage of time which a given walker 
spends in the backbone $B$ during the time interval $[0,t]$. 
We assume that the last relation applies also to the diffusion within
the periodic channels when the $H$ regions are long and narrow enough. 
Writing 
$$
t F_B(t) = \int_{0}^t\! du P_S(u), 
$$
where $P_S(t)$ in the probability to be in the Shaft region 
(corresponding to the backbone) at time $t$.
We extend the above equation to the continuous time case,    
\begin{equation}
\frac{1}{2D_{0}}\frac{d\langle [x(t)-x(0)]^{2}\rangle}{dt} = 
P_{\mathrm{S}}(t) 
\label{eq:generalizedDx2dt}
\end{equation}
from which, by an integration, we recover the expression \eqref{eq:Dt}. 

The above reasoning can be repeated in the presence of 
an imbalance in the jump probabilities
along the backbone, i.e. by considering that the random variable 
$\xi_{j}$ takes values in $\{1,0,-1\}$, respectively with probability 
$\{1/4-\varepsilon,1/2,1/4+\varepsilon\}$. 
Thus a given walker experiences an external drift $f=2\varepsilon$, 
with $\varepsilon \in [0,1/4)$. Via a similar calculation used to derive 
$\langle[x(t)-x(0)]^{2}\rangle_{0}$, we get
\begin{equation}
\langle \Delta x(t)\rangle_{f} = f t F_{\mathrm{B}}(t);
\label{eq:drift_backbone}
\end{equation}
thus, on the comb lattice, we easily obtain
\begin{equation}
\frac{\langle\Delta x(t)\rangle_{f}}{\langle \Delta x^2(t) \rangle_{0}}=
\frac{f}{2}
\label{eq:fdt_app}
\end{equation}
The last result which is exact for the comb lattice, however, 
can be a guide for the interpretation of diffusion and linear 
response within periodic channels, as shown in Fig.~\ref{fig:Response}.

Let us conclude by noting that the relation \eqref{eq:fdt_app} not only remains true 
for any time, i.e. in asymptotic and pre-asymptotic regimes, but also for a
generic initial distribution of walkers. Such a conclusion is obvious 
by observing that Eqs.\eqref{eq:msd_backbone} and \eqref{eq:drift_backbone} 
are both proportional to the same function $F_{\mathrm{B}}(t)$ which does 
depend on the initial distribution but simplifies in the ratio 
\eqref{eq:fdt_app}.

\end{document}